\RequirePackage{fix-cm}
\documentclass[smallextended]{svjour3}       
\smartqed  
\usepackage{graphicx}
\begin{document}

\title{Absence of superfluidity in 2D dipolar Bose striped crystals
}


\author{Fabio Cinti         \and
        Massimo Boninsegni 
}


\institute{F. Cinti \at
              Department of Mathematics and Applied Mathematics, University of Johannesburg, Johannesburg, South Africa\\
              \email{fcinti@uj.ac.za}           
           \and
           M. Boninsegni \at
              Department of Physics, University of Alberta, Edmonton, Alberta, Canada, T6G 2E1\\
              \email{m.boninsegni@ualberta.ca}
}

\date{Received: date / Accepted: date}

\maketitle

\begin{abstract}
We present results of computer simulations at low temperature of a two-dimensional system of dipolar bosons, with dipole moments aligned at an arbitrary angle with respect to the direction perpendicular to the plane. The phase diagram includes a homogeneous superfluid phase, as well as triangular and striped crystalline phases, as the particle density and the tilt angle are varied. In the striped solid,  no phase coherence among stripes and consequently no ``supersolid" phase is found, in disagreement with recent theoretical predictions.
\keywords{Supersolid phase \and Dipolar systems \and Quantum Monte Carlo}
\end{abstract}

\section{Introduction}
\label{intro}
Quantum many-body systems in which the elementary constituents possess a finite dipole moment have elicited significant experimental and theoretical interest, motivated by speculations of novel exotic phases of matter that the long-ranged, anisotropic character of the interaction may underlie (see, for instance, Ref. \cite{menotti}).  One such phase is the supersolid \cite{bp12}, namely a homogeneous phase of matter which displays crystalline order and is capable at the same time of sustaining  flow without dissipation.
\\ \indent
After a few decades of intense but so far unsuccessful search for evidence of supersolid behavior in $^4$He, attention has turned to cold atoms, which offer the advantage of a highly controllable physical setting with tunable inter-particle interactions \cite{feshbach}. Indeed, theoretical proposals have been made of specific cold atom systems allowing in principle for the observation of supersolid behavior, for example with Rydberg atoms \cite{cinti10,saccani1,saccani2,jltp,maucher,nc}; it should be also noted that experimental evidence of novel phases simultaneously displaying density ordering and superfluidity has been  reported for atomic Bose-Einstein Condensates (BECs) with spin-orbit interactions \cite{macche1}, or coupled to the modes of optical cavities \cite{macche2}. More recently, the experimental investigation of a possible three-dimensional, ``droplet" supersolid phase of dipolar bosons with aligned dipole moments \cite{ferrier,chomaz}, which has been theoretically predicted \cite{cb17,ccsm17,mc19}, has started to yield promising results \cite{tanzi,bottcher,chomaz2}. 
\\ \indent
Dipolar Bose systems were first suggested as a viable candidate for a supersolid phase in the (quasi)-2D limit, with dipole moments all aligned in the direction perpendicular to the plane, in which case the inter-particle interaction is purely repulsive. A supersolid phase was hypothesized in the form of an ordered lattice of large solid clusters floating  in a superfluid background, occurring in lieu of the conventional coexistence of crystal and fluid phases separated by a macroscopic interface \cite{spivak}. First principle calculations of the width of the coexistence region, as well as of the energy of the interface, have however ruled out such an intriguing scenario, at least for practical purposes \cite{bm14}. 
\\ \indent 
It has been subsequently proposed that a superfluid crystal in this system might be underlain by three-body interactions \cite{gora}, or might alternatively arise by tilting the aligned dipoles with respect to the axis perpendicular to the physical plane in which particles are confined, leading to the appearance of a striped phase \cite{baranov}. The claim was recently made that such a phase displays ``supersolid" behavior, in a range of density and tilt angle \cite{soliti}.  
Although the supersolid scenario has been investigated experimentally in different recent works \cite{tanzi,bottcher,chomaz2,wenzel}, no close realization of the system in the quasi-two-dimensional limit explored in Ref. \cite{soliti} has yet been achieved; nevertheless, it remains of fundamental theoretical interest, and one for which, at least in principle, a direct comparison of theory and experiment may be possible at some point in the future.
\\ \indent
In this paper, we present results of a first principle numerical study of the phase diagram of dipolar bosons in 2D, with their dipole moments aligned at an arbitrary angle with respect to the direction perpendicular to the plane. Our theoretical model of the system is identical with that of Ref. \cite{soliti}, but we make use of a different numerical technique, a finite temperature one which, unlike those adopted in Ref. \cite{soliti}, does not rely on any {\em a priori} physical assumption on the physics of the system, and allows for an unbiased calculation of all cogent physical observables. This turns out to be a key point, as we shall see.
Our computed phase diagram is  in agreement with that of Ref. \cite{soliti} {\em only} as far as the structure of the various phases is concerned, namely triangular crystal, striped crystal and superfluid. On the other hand, our results show {\em no evidence at all} of  the superfluid behavior of the striped crystal proposed in Ref. \cite{soliti}. In particular, we find the superfluid response of the system in the striped solid phase  to vanish in the thermodynamic limit, including for values of the density and/or tilt angle  for which a ground state superfluid response $\sim$100\% is reported in Ref. \cite{soliti}. We attribute this radical disagreement between our predictions 
and those of Ref. \cite{soliti}, to the computational methodology utilized therein, which is intrinsically biased,  and has in the past yielded several spurious, eventually   disproved indications of superfluidity in other physical systems.  
\\ \indent
The remainder of this article is organized as follows: in Section \ref{meth} we describe the model of the physical system and the computational methodology adopted in this study; in Section \ref{res} we present our results, while in Section \ref{concl} we outline our physical conclusions and discuss the disagreement between our results and those of Ref. \cite{soliti}.
\section{Methodology}\label{meth}
We consider an ensemble of $N$ Bose particles of spin zero,  mass $m$ and dipole moment $D$,  moving in 2D. All dipole moments are aligned, pointing in an arbitrary direction forming an angle $\alpha$ with the axis ($z$) perpendicular to the plane.
Henceforth, we express all lengths in terms of the characteristic length of the dipolar interaction, namely $a\equiv mD^2/\hbar^2$,
whereas $\epsilon\equiv (D^2/a^3)=\hbar^2/(ma^2)$ is the unit of energy and temperature (i.e., we set the Boltzmann constant $k_B=1$).
The  Hamiltonian of the system in dimensionless
units reads as follows:
\begin{equation}\label{u}
\hat H =-\frac{1}{2}\sum_i\nabla^2_i+\sum_{i<j}U({\bf r}_i,{\bf r}_j)
\end{equation}
where ${\bf r}_i\equiv (x_i,y_i)$ is the position of the $i$th particle in the plane,
and the interaction  $U$ between any two particles is given by the classical dipolar potential, namely
\begin{equation}\label{dipo}
U({\bf r},{\bf r}^\prime)=\frac{1}{|{\bf r}-{\bf r}^\prime|^3}\ \biggl (1-3\   {\rm cos}^2\theta\ {\rm sin}^2\alpha\biggr )
\end{equation}
where $\theta$ is the angle between the two vectors {\bf r} and {\bf r}$^\prime$.
The system is enclosed in a rectangular cell with periodic boundary conditions. At temperature $T=0$ the only two thermodynamic parameters are the system density $n$ and the angle $\alpha$.
\\ \indent
A system of particles interacting through (\ref {dipo})  is thermodynamically stable only for a tilt angle $\alpha \le \alpha_c={\rm sin}^{-1}(1/\sqrt 3)$, as the dipolar interaction features an infinitely deep attractive well for greater $\alpha$, causing the system to collapse unless a short-range repulsive part is added to (\ref{dipo}), a repulsion that is always present in real physical systems. The ground state phase diagram of this model for $\alpha=0$ only includes a crystalline (triangular) and a superfluid phase \cite{bm14}.
For $0 < \alpha< \alpha_c$, the anisotropy of the interaction promotes the formation of stripes, and the theoretical question is whether a phase featuring both crystalline (striped) order and superfluidity is possible.
\\ \indent
In order to address this issue we have carried out extensive Quantum Monte Carlo simulations of the system as function of $\alpha, n$, based 
on the continuous-space Worm Algorithm \cite{worm,worm2}, specifically a variant thereof in which the number of particles $N$ is fixed \cite{fabio,fabio2}. Our simulations are at finite temperature; because we are ultimately interested in the physics of the ground state, we need to extrapolate the results to the $T\to 0$ limit.
\\ \indent
Details of the simulation are  standard. We used  the primitive approximation for the short imaginary time ($\tau$)  propagator, and report here numerical estimates for structural and superfluid properties of interest extrapolated to the $\tau\to 0$ limit. In general, physical estimates obtained with a value of the time step $\tau\sim 1.5\times 10^{-5}\ \epsilon^{-1}$ are indistinguishable from the extrapolated ones, within the statistical uncertainties of the calculation.
We carried out simulations of systems comprising a variable number of particles, 400 being the largest size utilized in this work; as we shall see, however, the most important physical conclusions can be established on systems of smaller size. We typically started our simulations from many-particle configurations corresponding to the classical ground states for the chosen values of $n$ and $\alpha$. However, we also carried out simulations with a different initial configurations, and verified convergence to the same physical results.
\\ \indent
The superfluid properties of the system are the most important aspect of this work. We computed the superfluid fraction  $\rho_S$ directly, by means of the well-known ``winding number" estimator \cite{pollock}; we also assessed the presence of off-diagonal quasi-long-range order, which characterizes a superfluid transition in 2D, by computing the one-body density matrix $n(r)$, and studying its behavior as a function of temperature. 
\section{Results}\label{res}
\begin{figure}
\centering
\includegraphics[width=3in]{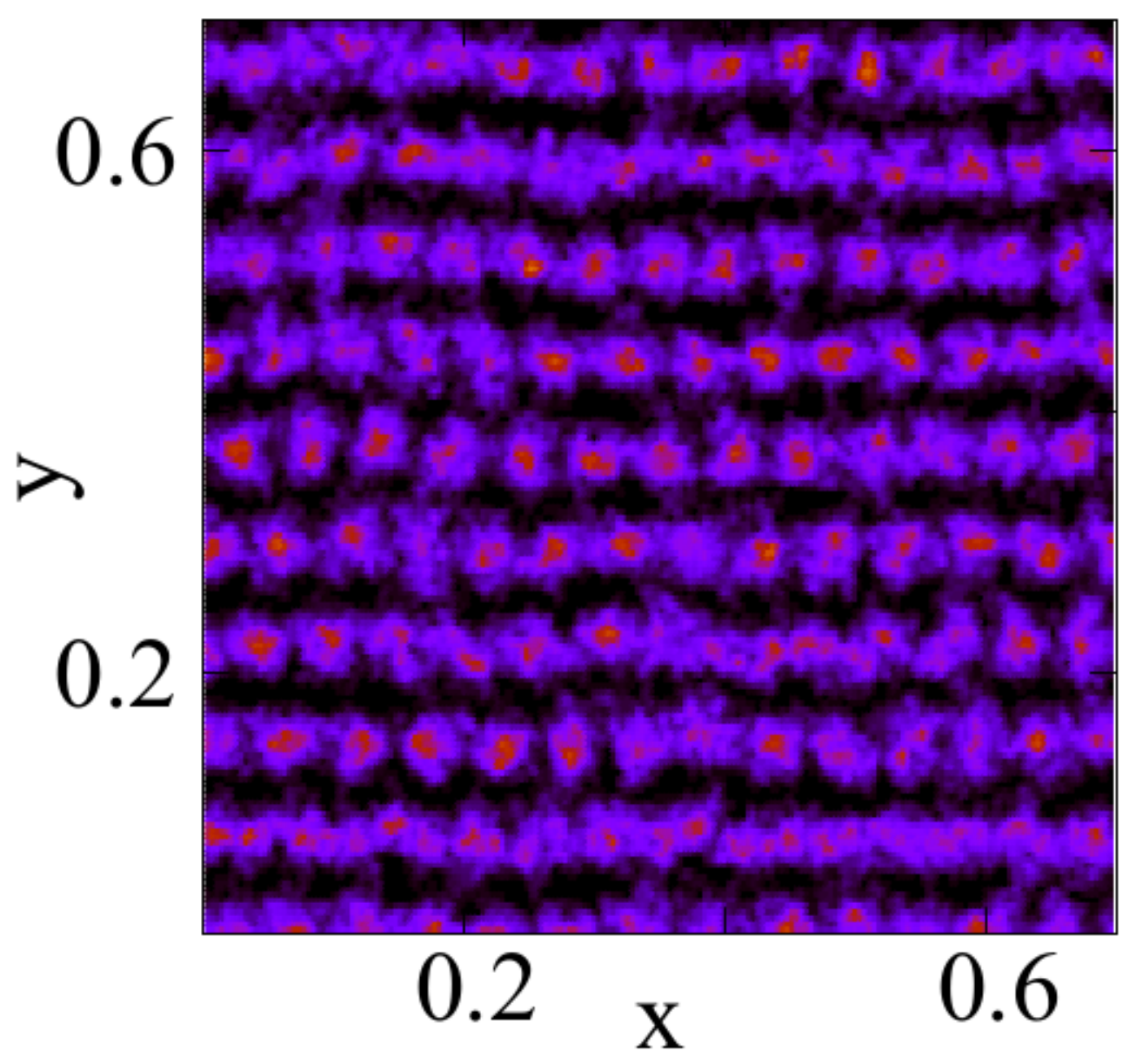}
\caption{{\it Color online}. Density map for a system of 2D dipolar bosons with aligned dipole moments in a direction making an angle $\alpha =0.55$ rads with respect to the direction perpendicular to the plane. The density is $n=256$ and the temperature $T=60$, in the units utilized here. 
}\label{f1}
\end{figure}
As stated above, the low temperature phase diagram obtained in this work is in agreement with that of Ref. \cite{soliti} (Fig. 1 therein), as far as the structure of the various phases is concerned. Specifically, a triangular crystal, a homogeneous superfluid, and a striped crystalline phases are observed, and the phase boundaries are in quantitative agreement with those of Ref. \cite{soliti}, although it is important to note that the crystalline ground states, both striped as well as triangular, are always found to be {\em commensurate}, i.e., free of point defects such as vacancies and interstitials, a fact that is known to have implications on the superfluid response \cite{prosvi,dang}. 
\\ \indent
There is, however, complete disagreement when it comes to the characterization of the (putative) superfluid properties of the striped phase, to which we restrict our discussion in the remainder of this paper. In particular, according to Ref. \cite {soliti} the striped crystal {\em always} displays a finite superfluid response, generally anisotropic and typically with a rather small value of the superfluid fraction in the direction perpendicular to the stripes (taken here to be the $y$ direction); however,
the superfluid fraction reported in Ref. \cite{soliti} is essentially 100\% in {\em both} directions  for three  specific thermodynamic $(\alpha,n)$ points, namely $K\equiv (0.6,128)$, $I\equiv(0.55,256)$ and $F\equiv(0.5,400)$ (angles are in rads)  \cite{strange}.
\\ \indent
Figure \ref{f1} shows a  two-dimensional density map for a typical striped phase corresponding to point $I$ defined above; the temperature $T=60$. The arrangement of particles in parallel stripes is clear \cite{mapexplain}.
Because the system is anisotropic, one generally expects $\rho_S^x\ne\rho_S^y$, but {\em both} must be finite in the low temperature limit, in a genuine 2D supersolid; henceforth, we focus for simplicity on the response $\rho_S^y$ in the transverse direction. 
At these physical conditions, we find $\rho_S^y$ to amount to statistical noise, i.e., zero within the combined statistical and systematic uncertainties (which we estimate $< 10^{-3}$), with no detectable change as $T$ is lowered from $T=240$ to $T=60$. It should be mentioned that this is observed on a relatively small (168 particles) system, for which a spurious superfluid signal could be expected, and despite the occurrence of many-particle exchanges of significant length (see below).
Obviously, the question immediately arises of how low $T=60$ is, i.e., whether a hypothetical superfluid transition in this system might take place at significantly {\em lower } temperature.
On this point, we note that a superfluid transition in 2D must conform to the Kosterlitz-Thouless (KT) paradigm, and therefore satisfy the universal jump condition \cite{kt} at the critical temperature $T_c$, which in turn implies, assuming $\rho_S (T=0)\approx 1$ as reported in Ref. \cite{soliti}, that $T_c$ should be $\approx T^\star = n = 256$ (see, for instance, Ref. \cite{filinov}). Thus, at $T=60$ (i.e., $\sim T^\star/4)$ a robust superfluid response should be observed, were the predictions of Ref. \cite {soliti} correct. 
\begin{figure}[t]
\centering
\includegraphics[width=3.2in]{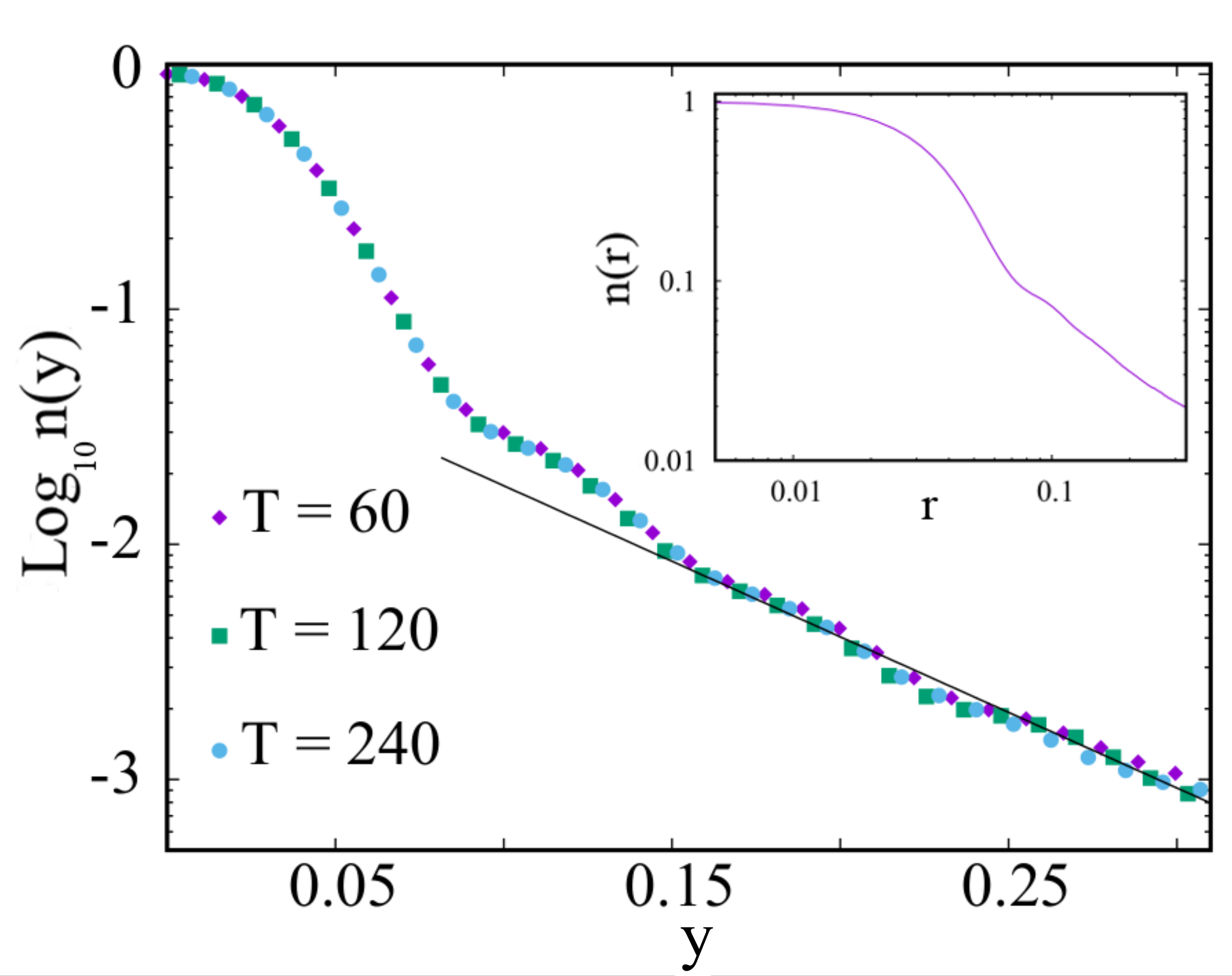}
\caption{{\it Color online}. One-particle density matrix $n(y)$, evaluated along the direction perpendicular to the stripes, for the system at density $n=256$ and with angle $\alpha=0.55$ rads, at three different temperatures, namely $T=240, 120$ and 60 in the units used here. Statistical errors are smaller than the symbol sizes. Solid line is an exponential fit to the long distance part of the function. Inset shows for comparison the one-body density matrix for the system at the same density and at temperature $T=240$  (note the log-log scale), but with tilt angle $\alpha=0.4$ rads. The system is in this case a homogeneous superfluid ($\rho_S\approx 0.8$ at this temperature).} 
\label{f2}
\end{figure}
\begin{figure}[h]
\centering
\includegraphics[width=3.2in]{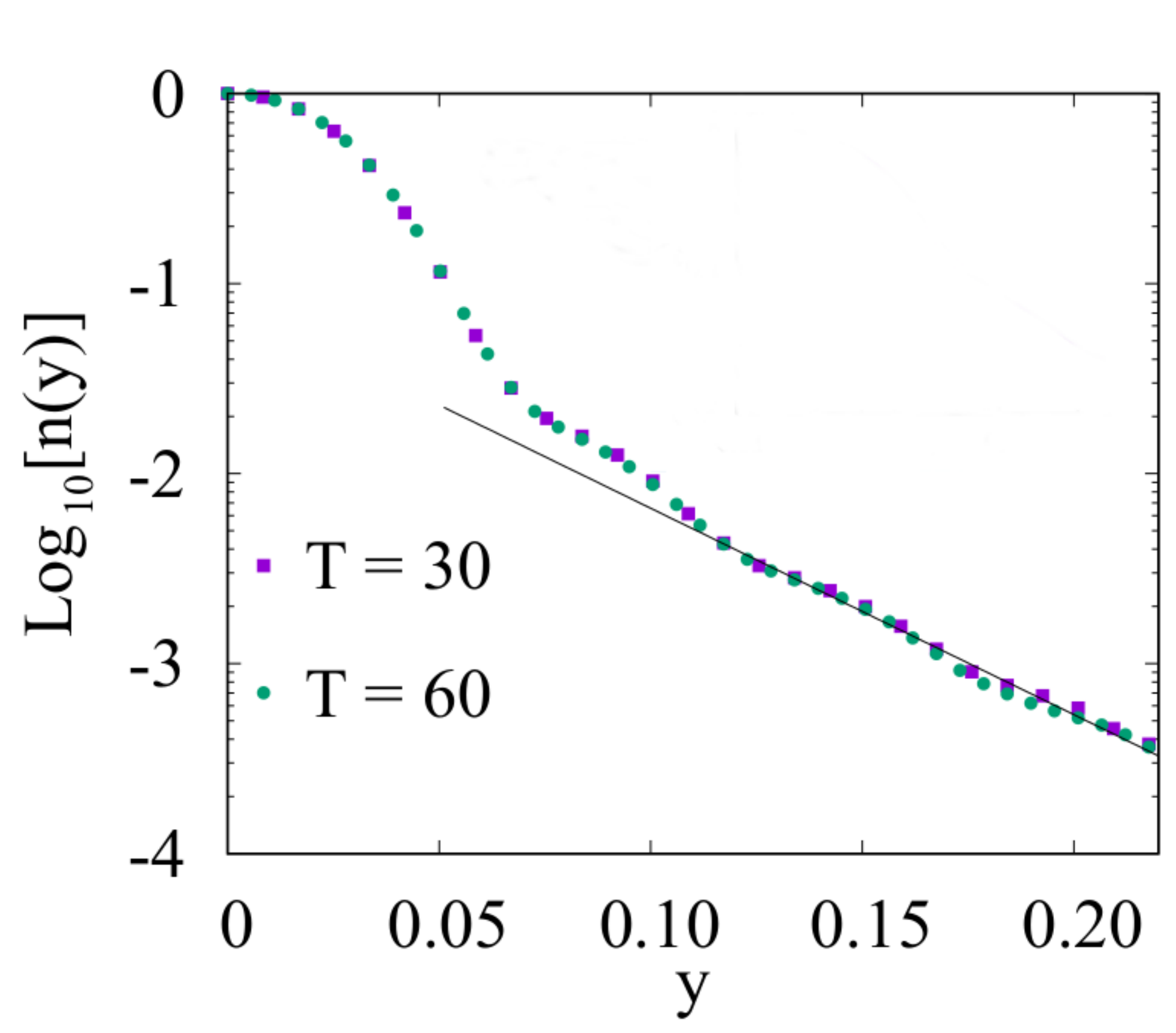}
\caption{{\em Color online}. Same as Fig. \ref{f2} but at density $n=400$ and temperature $T=60, 30$.}
\label{f3}
\end{figure}
\\ \indent 
Fig. \ref{f2} shows the one-body density matrix $n(y)$, computed along the transverse direction at density $n=256$, tilt angle $\alpha=0.55$ and at the three temperatures $T=240, 120 $ and 60, in the units utilized here. The long-distance behavior is clearly suggestive of, and quantitatively consistent with exponential decay, which is observed up to a distance $\sim 5$ times the interparticle distance (which is actually comparable to the distance between stripes, see Fig. \ref{f1}). In this interval, $n(r)$ decays by three orders of magnitude. No discernible dependence of the results on the temperature can be observed, within the statistical error of the calculation, which is consistent with the behavior of a non-superfluid insulator (see, for instance, Ref. \cite{bp12}).
\\ \indent 
\begin{figure}[h]
\centering
\includegraphics[width=3.2in]{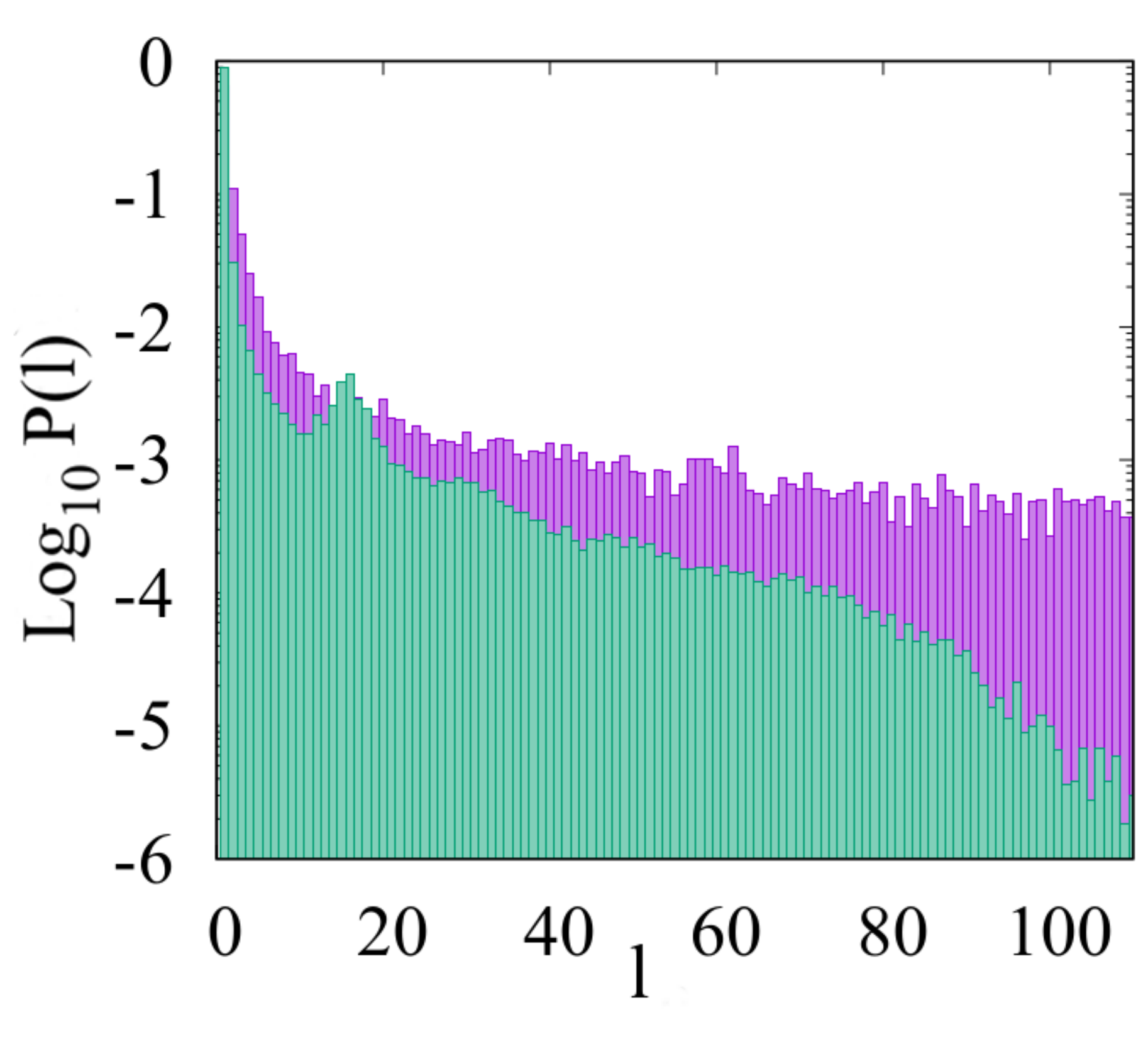}
\caption{{\it Color online} Frequency of occurrence of exchange cycles involving $l$ particles for two systems with density $n=128$ at temperature $T=30$. Darker boxes refer to the case $\alpha=0.5$ rads, for which the system is 100\% superfluid, whereas lighter ones to $\alpha=0.6$, for which the system is in the striped crystal phase and the transverse superfluid response  $\rho_S^y< 0.001$. Note the logarithmic scale on the $y$-axis.
}\label{f4}
\end{figure}
This can be contrasted with the behavior shown in the inset of Fig. \ref{f2}, of the (circularly averaged) one-body density matrix $n(r)$ for a system at the same density at at temperature $T=240$, but with tilt angle $\alpha=0.4$ rads. The system in this case is in the superfluid phase, i.e., no crystalline order is present, and the computed superfluid fraction is $\rho_S = 0.8$, within statistical uncertainties. Here, $n(r)$ displays the characteristic, slow power law decay of a 2D superfluid.  Based on these results, we conclude that $\rho_S^y=0$ in the ground state of the system, if $\alpha=0.55$ rads.
\\ \indent
Fig. \ref{f3}  shows the one-body density matrix for the thermodynamic point $F$ defined above. Here too, we have explicitly observed the same physical behavior described above, namely absence of any measurable superfluid response and exponentially decaying one-body density matrix in the transverse direction, at temperatures as low as $T=30$, i.e., 0.075 $T^\star$. The same conclusion applies  to the other  thermodynamic point defined above (i.e., $K$), and {\em a fortiori} to the rest of the phase diagram of the system, in the region which a striped crystal occurs, as we have verified by performing a few targeted simulations.
\\ \indent
The markedly different superfluid response observed for different values of the tilt angle $\alpha$ at the same density and temperature is reflected in the frequency of occurrence of cycles of exchanges involving $l$ particles, $P(l)$, an example of which is shown in Fig. \ref{f4}. Here, the density $n=128$ and $T=30$; results for the two different angles $\alpha=0.5$, for which the system is a fluid with $\rho_S^y=1$, and $\alpha=0.6$, for which the system is in the striped crystal and $\rho_S^y=0$ (in both cases the quoted values are within statistical uncertainties). Although exchanges involving significant numbers of particles occur in both cases, for $\alpha=0.6$ they remain mostly local in character, the frequency of longer cycles decaying very rapidly with $l$. 
Quite generally, the striped crystal behaves essentially as a collection of parallel, largely independent one-dimensional chains. Although the (quasi)superfluid properties of these chains (in the Luttinger sense) could be of interest \cite{stan}, in no case can this system be meaningfully regarded as a ``supersolid".  Thus, allowing for a finite ``tilt" angle only leads to an additional (striped) crystalline phase, with respect to the extensively investigated  $\alpha=0$ case, but does not lead  to novel superfluid behavior in this purely 2D system. The third ($z$) dimension is crucial in order to stabilize a supersolid phase.
\section {Discussion}\label{concl}
It is clearly necessary to assess the origin of such a major, quantitative and qualitative disagreement between our calculation and that of Ref. \cite{soliti}, as both are based on the same microscopic Hamiltonian and employ numerical techniques which should yield compatible results, within statistical errors. 
The calculations of Ref. \cite {soliti} are based on two different ground state techniques, namely Diffusion Monte Carlo (DMC) (see, for instance, Ref. \cite{dimc}) and Path Integral Ground State (PIGS) \cite{sarsa,cuervo}. The disagreement between their results and the ones offered here, as mentioned above, is in the estimates of the superfluid fraction, as well as in the long-range behavior of the one-body density matrix, which is claimed in Ref. \cite{soliti} to display a slow, power-law decay at long distances, as opposed to the exponential decay found here.
 \\ \indent 
As mentioned above, the phase boundaries obtained in this work are in agreement with those of Ref. \cite{soliti}. In a $T=0$ calculation, phase boundaries are typically established through a comparison of the energetics arrived at by projecting the lowest-energy state out of initial trial wave functions featuring different kinds of order, and assuming that the equilibrium phase is that of the wave function yielding the lowest energy estimate. Although a direct comparison of the energetics obtained in this work and in Ref. \cite{soliti}  cannot be carried out, as energy values are not furnished therein, it is altogether not surprising that there is agreement between DMC/PIGS and this work. For, it is mainly in the calculation of expectation values of observables that do not commute with the Hamiltonian, chiefly those associated wit the superfluid response, that significant differences arise between the results obtained with ground state and finite temperature methods, due to the fact that DMC/PIGS estimators for the relevant quantities (e.g., the one-body density matrix) are inherently biased (we come back to this point below). 
It need be emphasized again, however, that in this work crystalline ground states are always found to be {\em commensurate}, i.e., free of  point defects like vacancies or interstitials. This is a potentially very important point because, while it is possible for a crystalline system to feature a nonzero superfluid response in the presence of such point defects, nevertheless such a phase is thermodynamically unstable \cite{fate} with respect to one that is commensurate, i.e., free of defects, for which the superfluid response vanishes \cite{mlc}; this aspect may have been overlooked in Ref. \cite {soliti}, as it is not discussed at all.\\ \indent
The second point that has to be made is that,
although often advertised as ``exact'', ground state methods are in fact affected by an inherent bias, associated to the trial wave function out of which the ground state is projected \cite{notec}. Such bias (which does {\em not} affect finite temperature techniques, as they  require no {\em a priori} input),  is often impractically 
difficult to remove (even with very long computer runs \cite{sm,psb}); moreover, for the some cogent quantities it is in fact {\em not removable at all}, at least not in any systematic way \cite{dimc}. For example, there exists no numerically exact, unbiased procedure to compute $n(r)$ within either DMC or PIGS \cite{noteob}; one has to resort to the so-called ``mixed estimators", which by construction depend on the trial wave function utilized  (see, for instance, Ref. \cite{entropy}).\\ \indent
In other words, the results for $\rho_S$ and $n(r)$ offered in Ref. \cite{soliti} are intrinsically only approximate, and affected by a {\em systematic} error due to the specific choice of trial wave function, an error that is ultimately not even quantifiable.  In light of all that,  the contentions made in Ref. \cite{soliti} of, e.g., power-law decay of $n(r)$   at long distance in the striped crystal phase, much less estimates of the condensate fraction of, e.g, $\sim 10^{-3}$ with a claimed relative precision of 10\%, made using either PIGS or DMC are not at all believable, especially if accompanied by estimates of the superfluid fraction unphysically close to unity, for a system breaking translational invariance.
It is worth mentioning the numerous past predictions of superfluidity of various Bose systems, made using ground state techniques, which were subsequently 
proven incorrect \cite{11,13,14,15,17,ultimo}. 
Indeed, finite temperature techniques are now widely regarded as a far superior option for investigating the ground state of Bose systems (for an extensive discussion  of this subject, see for instance Ref. \cite{rmp95}). 
\section*{Acknowledgments}
This work was supported in part by the Natural Sciences and Engineering Research Council of Canada (NSERC). Computing support from ComputeCanada is gratefully acknowledged.

\section*{Conflict of interest}
The authors declare that they have no conflict of interest.


\begin{thebibliography}{99}
\bibitem{mapexplain}
The map shown in Fig. \ref{f1} is obtained from an instantaneous configuration (particle world lines) generated in the course of a long simulation. It is physically equivalent to all other configurations generated in the same run (e.g., others differ at the most by an overall rigid translation). It can be therefore regarded as representative of the density of the system.
\bibitem{notec}
Additional sources of bias in ground state methods are the necessarily finite projection time and/or the population of random walkers utilized.
\end{thebibliography}
\end{document}